\documentclass[a4paper,11pt]{article}%
\usepackage{geometry}
\usepackage{amsmath}
\usepackage{amsfonts}
\usepackage{amssymb}
\usepackage{graphicx}
\usepackage{indentfirst}
\usepackage[small,bf]{caption}
\usepackage{slashed}
\usepackage[normalem]{ulem}
\usepackage{xcolor}
\usepackage{ulem}
\providecommand{\U}[1]{\protect\rule{.1in}{.1in}}

\renewcommand{\d}{\ensuremath{\mathrm{d}}}

\newcommand{\MSbar}{\overline{\mbox{MS}}}

\newcommand{\p}{\partial}
\newcommand{\tr}{\ensuremath{\mathrm{Tr}}}
\newcommand{\e}{\ensuremath{\mathrm{e}}}

\newcommand\redsout{\bgroup\markoverwith{\textcolor{red}{\rule[0.5ex]{2pt}{0.8pt
}}}\ULon}

\def\bs{b\!\!\!/}

\def\ds{\partial\!\!\!/}

\DeclareMathOperator{\Tr}{Tr}

\usepackage{color}
\definecolor{darkgreen}{rgb}{0,0.35,0}
\definecolor{Rood}{rgb}{1, 0, 0}

\newcommand{\mx}{\ensuremath{\mathsf}}

\def\XXint#1#2#3{{\setbox0=\hbox{$#1{#2#3}{\int}$}\vcenter{\hbox{$#2#3$}}\kern-.5\wd0}} 

\begin{document}

\date{}
\title{\textbf{Non-Abelian extension of the aether term and the Gribov problem}}
\author{
\textbf{D.~R.~Granado}$^{a,b}$\thanks{diegorochagranado@duytan.edu.vn}\,\,,
\textbf{C.~P.~Felix}$^{c}$\thanks{carolfba@gmail.com}\,\,,
\textbf{I.~F.~Justo}$^{d}$\thanks{igorfjusto@gmail.com}\,\,,
\textbf{A. Yu. Petrov}$^{e}$\thanks{petrov@fisica.ufpb.br}\,\,,
\textbf{D. Vercauteren}$^{a,b}$\thanks{vercauterendavid@duytan.edu.vn}\\
{\small \textnormal{$^{a}$  \it Institute of Research and Development,}} \\ 
{\small \textnormal{ \it Duy T{a}n University, D{a} Nang 550000, Vietnam}\normalsize} \\ 
{\small \textnormal{$^{b}$  \it Faculty of Natural Sciences,}} \\ 
{\small \textnormal{ \it Duy T{a}n University, D{a} Nang 550000, Vietnam}} \\ 
{\small \textnormal{$^{c}$ \it Department of Physics, Chung Yuan Christian University (CYCU),}}\\
{\small \textnormal{\it No. 200, Zhongbei Road, Zhongli District, Taoyuan City, 320, Taiwan.}}\\
{\small \textnormal{$^{d}$ \it Universidade Federal Fluminense, Instituto de F\'isica,}} \\
{\small \textnormal{\phantom{$^{c}$} \it Av. Litoranea s/n, 24210-346, Niter\'oi, RJ, Brasil}} \\
{\small \textnormal{$^{e}$  \it Departamento de F\'{i}sica, Universidade Federal da Para\'{i}ba,}}\\
 {\small \textnormal{ \it Caixa Postal 5008, 58051-970, Jo\~{a}o Pessoa, 
Para\'{i}ba, Brazil}\normalsize}}
\maketitle

\begin{abstract}
In this paper, we treat the proper path integral quantization
of the Yang-Mills-aether (YM-aether) system by dealing with the extra gauge copies {in} the Landau gauge.
Within Gribov's prescription to get rid of such remaining gauge copies, we
explicitly derive the Gribov parameter dependence of the coupling constant and
of the Lorentz violation aether term. The ultraviolet limit is investigated
under the light of recent bounds on the magnitude of the non-Abelian aether parameter, and we show that the Gribov parameter can be disregarded in that
limit.
\end{abstract}

\section{Introduction}

The possibility of Lorentz symmetry breaking (LSB) has recently been discussed 
in many different contexts. It was proposed for the first time in the context of QED by Carroll, Field and Jackiw (CFJ) in 90s \cite{CFJ}: they suggested
a consistent Lorentz-breaking extension of a known quantum field theory
model involving a constant axial vector $b_{\mu}$. This axial vector induces a
privileged space-time direction, therefore breaking the Lorentz symmetry. Soon after that, a variety of Lorentz-breaking extensions of the standard model were put forward for consideration by others
\cite{ColKost} and many nontrivial issues related with
these proposals have been 
discussed. Among these issues, we can emphasize an unusual wave propagation, which includes 
birefringence and rotation of the polarization plane of an electromagnetic field in
vacuum (cf. \cite{Ja}) that has been shown to take place in various 
Lorentz-breaking extensions of QED (cf. \cite{MP,Casana}); ambiguities in the quantum 
corrections (cf. \cite{Ja1}); and perturbative generation of new Lorentz-breaking 
terms (cf. \cite{CFJ}). Plenty of experimental measurements of potential signals of Lorentz symmetry breaking have 
been carried out in different cases (\textit{Cf.}
\cite{Kostelecky:2008ts} and references therein). Also, the 
renormalizability of Lorentz violating QED was discussed in 
\cite{Santos:2016bqc}.

As was previously mentioned, Lorentz symmetry breaking was treated in the QED context. Naturally, one can ask for a non-Abelian 
extension of the Lorentz-breaking terms. (A list of possible Lorentz-breaking additive terms, including non-Abelian ones is given in \cite{KosLi}.) The non-Abelian Carrol-Field-Jackiw (CFJ) term can be
generated perturbatively (see \cite{ourYM} for more details) and some consequences when adding 
this term were also discussed \cite{Santos:2016dcw,Santos:2016uds,Santos:2014lfa}.
The first systematic analysis of such a theory 
was presented in \cite{ColMac}. The renormalizability of some non-Abelian systems involving Lorentz symmetry 
breaking has been explored as well
\cite{Santos:2016dcw,Santos:2016uds,Santos:2014lfa}. Recently, the authors of \cite{Granado:2017xjs} 
studied the path integral quantization of the YM+CFJ system.

It is appealing to consider whether a non-Abelian generalization 
of other Lorentz-breaking terms is possible as was done for the CFJ case. The intuitive candidate for such a 
generalization is the aether term \footnote{Name originally adopted in \cite{Carroll} (where -- and only in that paper -- a certain reminiscence with a pre-Einsteinian aether was really noted) and further used in \cite{Gomes:2009ch,aether2}.}, 
which, unlike the CFJ term, does not break the CPT symmetry. The classical aspects 
of this term, together with its possible implications within the framework of extra 
dimensions, were intensively discussed in \cite{Carroll}. The Abelian 
perturbative generation of the aether term has been carried out for the first 
time in \cite{Gomes:2009ch}, and in \cite{aether2} the aether term was shown to 
be strongly ambiguous. The non-Abelian generalization of the aether 
term looking like ${\cal L}^{(4)}_A=-\frac{1}{4}k^{(4)\kappa\lambda\mu\nu}{\rm tr}(F_{\kappa\lambda}F_{\mu\nu})$ has been proposed \cite{KostLi}. Afterwards, it has been shown in 
the paper \cite{ournAb}, that this term 
arose as a one-loop correction in a theory involving coupling of a non-Abelian gauge field with spinor matter.
Explicitly, the starting point is the following functional trace
\begin{equation}
{\Gamma^{(1)}_{\mathrm{min}}}=i{\rm 
Tr}\ln(i\delta_{ij}\ds-e\gamma^{\mu}A^a_{\mu}(T^a)_{ij}-m\delta_{ij}-\bs\gamma_5\delta^{ij}).
\end{equation}
In \cite{ournAb}, it was found that the aether-like result, that is, the contribution of second order in the constant axial vector $b_{\mu}$ and up to second order in derivatives of the gauge field $A_{\mu}^a${,} is
\begin{equation}
\label{aether}
\Gamma^{(1)}=-\kappa\frac{e^2}{6\pi^2m^2}b^{\mu}F_{\mu\nu}^ab_{\lambda}F^{\lambda\nu a},
\end{equation}
where $F^{\mu\nu a}=\partial^{\mu}A^{\nu a}-\partial^{\nu}A^{\mu a}-ef^{abc}A^{\mu a}A^{\nu c}$ is the non-Abelian stress tensor, and
the constant $\kappa$ is defined from the trace condition ${\rm tr}(T^aT^b)=\kappa\delta^{ab}$. The term (\ref{aether}) is a particular form of the term ${\cal L}^{(4)}_A=-\frac{1}{4}k^{(4)\kappa\lambda\mu\nu}{\rm tr}(F_{\kappa\lambda}F_{\mu\nu})$ \cite{KostLi} for the {specific choice} of the constant $k^{(4)\kappa\lambda\mu\nu}=C\frac{e^2}{m^2}(b^{\kappa}b^{\mu}\eta^{\lambda\nu}-b^{\kappa}b^{\nu}\eta^{\lambda\mu}-b^{\lambda}b^{\mu}\eta^{\kappa\nu}+b^{\mu}b^{\nu}\eta^{\kappa\lambda})$, with $C=\frac{2}{3\pi^2}$ is a number. Notice that, $F^{\mu\nu a}$ is not present from the very beginning, but it arises from the summation over two-, three- and four-point functions. Besides, the result is ambiguity-free. Certainly, it is important to
investigate the existence of an effective model  
taking the aether term into account.

However, in order to properly quantize a Lorentz symmetry breaking scheme in a
non-Abelian gauge theory, issues of gauge redundancy need to be addressed.
The standard Faddeev-Popov (FP) gauge fixing procedure is a well-known framework
to deal with such a problem. Nevertheless, in \cite{Gribov:1977wm}, Gribov
showed that the FP procedure is not enough to unambiguously fix the gauge
freedom of YM theories.
He demonstrated that, even after imposing the Landau (or Coulomb) gauge, there
still remain redundant gauge fields configurations, called Gribov copies; furthermore,
the existence of such a gauge fixing residual ambiguity is closely related to the
existence of FP operator zero-modes. Soon after Gribov's work, Singer showed that the Gribov problem is not an
inherited problem from a specific gauge, but of the gauge theory itself \cite{Singer:1978dk}. According to Singer, ``the Gribov ambiguity for the
Coulomb gauge will occur in all other (Lorentz covariant)
gauges''. In order to get rid of these ambiguities which remain after gauxe fixing, Gribov proposed to restrict the gauge field
path integral to a specific region, called the first Gribov region, where
the system is supposed to be free of infinitesimal gauge copies. Such a restriction is implemented by means of a Heaviside-step function and with a
consequent introduction of new mass parameter. Such a restriction of the path
integral must be made according to a \textit{no-pole} condition imposed
perturbatively on the ghost propagator. As a result the BRST symmetry is
(softly) broken\footnote{Very recently a BRST exact formulation of the
(R)GZ framework has been formulated and extended to the wider class of linear
covariant gauges,
\cite{Capri:2015ixa,Capri:2015nzw,Capri:2016aqq,Capri:2016aif,Capri:2016gut,Capri:2017bfd}.}; the gauge
field propagator is suppressed in the deep 
IR limit and has no asymptotic one-particle interpretation, according to the Osterwalder-Shrader criteria \cite{Osterwalder:1973dx}; 
and the ghost propagator has an IR enhanced ($\sim p^{-4}$) behavior,
which is not in agreement with most recent lattice data, though
\cite{Cucchieri:2007rg,Cucchieri:2009zt,Cucchieri:2010xr}. 

In 1989, Zwanziger further developed Gribov's original approach
\cite{Zwanziger:1988jt,Zwanziger:1989mf,Zwanziger:1992qr}. Zwanziger realized,
at that time, that Gribov's idea to get rid of the FP operator zero-modes
(those associated to infinitesimal gauge transformations) is, in fact, a
condition to be imposed on the lowest lying eigenvalues of the FP operator. As a
result, Zwanziger proposed an all order local action \cite{Capri:2012wx} in the Landau gauge,
whose functional integral is restricted to the first Gribov region (supposed
to be free of infinitesimal gauge copies). Such an improved approach has since then been known as Gribov-Zwanziger (GZ) approach. Just as in Gribov's
original approach, the GZ framework also leads to a drastic modification
of the gauge field propagator, such that it can no longer be interpreted as an
asymptotic physical particle. Besides, in the Gribov-Zwanziger approach the gauge
propagator is highly suppressed in the deep IR limit
\cite{Sorella:2009vt,Dudal:2009bf,Capri:2010hb,Dudal:2012sb,Lavrov:2011wb,Lavrov:2012gb}.
Inherent to the Gribov-Zwanziger approach, a mass parameter, called
the Gribov parameter, is consistently introduced in such a way that the gauge field
two-point function acquires complex conjugated poles. This excludes the
possibility of a K\"all\'en-Lehmann spectral representation
\cite{Sobreiro:2005ec,Vandersickel:2012tz}, since the propagator must be always
positive for that (see \cite{Hayashi:2018giz} for a recent study of the
connection between the existence of complex conjugate poles and positivity
violation of the K\"all\'en-Lehmann representation)\footnote{The gauge dependence
of the Gribov problem has been studied in \cite{Lavrov:2011wb,Lavrov:2013boa}.}.
In \cite{Osterwalder:1973dx}, Osterwalder and Shrader showed that a positivity
violation in the K\"all\'en-Lehmann spectral representation prevents
the propagator from having an asymptotic particle interpretation. In this sense,
Gribov proposed a ``confinement'' interpretation for the gauge field within his
framework.

Recently, a refined
approach to the Gribov framework, known as the Refined Gribov-Zwanziger (RGZ)
approach, considered the existence of (mass-)dimension 2 condensates (of the
gauge field and the Gribov ghosts) \cite{Vandersickel:2012tz,Dudal:2008sp}, which leads to a theoretical model in full
agreement with recent numerical data
\cite{Dudal:2010tf,Cucchieri:2011ig}. In this refined approach the gauge field
propagator still displays complex conjugated poles and is of Stingl type \cite{Stingl:1985hx,Stingl:1994nk}, with a resonable
agreement with lattice quantum field theory \cite{Dudal:2008sp,Dudal:2010tf,Cucchieri:2010xr,Cucchieri:2011ig}. A deeper
investigation of the geometrical properties of the Gribov issue within a lattice
framework was developed in \cite{Cucchieri:2012cb,Cucchieri:2013nja}. It is
important to mention that Dyson-Schwinger equation (DSE) technique have been
used for a long time in the study of non-local aspects of QCD
\cite{Buttner:1995hg,Tandy:1997qf,Roberts:2000aa}, and in particular they have been extensively applied to the investigation of the gauge field propagator
\cite{Alkofer:2000wg,Watson:2001yv,Alkofer:2003jj}. It is also remarkable that
since 1982 DSE approaches augmented with the so-called Pinch Technique are pointing to an infrared massive gluon propagator,
\cite{Cornwall:1981zr}, which
nowadays is in agreement with lattice data
\cite{Aguilar:2004sw,Aguilar:2006gr,Aguilar:2008xm,Strauss:2012dg,Maas:2011se}\footnote{\textit{Cf.} \cite{Huber:2018ned} for a recent nice  
review on the nonperturbative properties of Yang-Mills theories within the DSE
framework.}. Besides that, an alternative approach to Gribov's issue has been
recently developed \cite{Tissier:2010ts,Tissier:2011ey,Pelaez:2014mxa}.

In this work we will
investigate the effects of such a Lorentz violating term through the path integral
quantization procedure within the Landau gauge. The Gribov ambiguities will be
treated within the GZ framework, at first order in the loop expansion.

This paper is organized as follows: in Section 2, we review the Gribov-Zwanziger approach to the 
Gribov problem within the Landau gauge. In Section 3, we carry out the path 
integral quantization of the Yang-Mills-aether system in the Landau gauge and 
deal with the Gribov copies. Finally, in Section 4, we present a summary 
where the results and perspectives are
discussed.

\section{The Gribov-Zwanziger quantization procedure in the Landau gauge}
\label{gzquatization}

In this section, we provide a consistent introduction to the framework developed by 
Zwanziger \cite{Zwanziger:1988jt,Zwanziger:1989mf,Zwanziger:1992qr} ({\it Cf.} \cite{Dudal:2009bf,Vandersickel:2012tz} for a complete review {of} the GZ approach). Just as in {Gribov's} original proposal, there is a self-consistency condition known as the gap equation that 
must be satisfied also in Zwanziger's approach. It means that, according to Gribov-Zwanziger (GZ) formalism, the gap 
equation 
must be satisfied in order to consistently perform the path integral 
quantization of a non-Abelian gauge field theory in the Landau gauge (among other gauges).

As was said in the Introduction of this paper, Gribov showed that the Coulomb gauge is plagued 
by a gauge
fixing ambiguity \cite{Gribov:1977wm}. Originally, he proposed a mechanism to get rid of such ambiguities (known 
as Gribov copies),
\cite{Gribov:1977wm}. His proposal was to restrict the functional integration of 
the gauge field
to the region where the Faddeev-Popov (FP) operator is free of zero-modes, the 
so-called first
Gribov region. This restriction amounts to considering only gauge field 
configurations
corresponding to positive eigenvalues of the FP operator. Since the FP 
operator is closely
related to the ghost-anti-ghost two-point function, Gribov proposed to 
investigate the
influence of the gauge field on this function by computing it up to one loop. Now, we present the GZ formalism that fully implements the Gribov region into the path integral.

\subsection{The Gribov-Zwanziger framework}
\label{gribovsection}

In \cite{Zwanziger:1988jt,Zwanziger:1989mf,Zwanziger:1992qr}, Zwanziger showed how Gribov copies can be treated in {a} local and renormalizable way. Zwanziger realized that the Gribov region -- the restriction of the gauge field configuration space to the region where the FP operator is positive {definite} -- boils down to considering only gauge field configurations corresponding to the lowest (non-trivial) eigenvalue of the FP operator. Specifically, 
\[
Z ~=~ \int \mathcal{D}\phi \; \delta(\lambda_{min}[A]) \; \e^{-(S_{YM} + 
S_{gf})}
\;,
\]
where $S_{YM}$ stands for the action that contains the Yang-Mills terms, $S_{gf}$ is the action that contains the terms inherited from the gauge fixing procedure, $\lambda_{min}[A]$ accounts for the trace over the matrix of all the 
lowest-lying eigenvalues of the FP operator and $\phi$ accounts for all quantum fields involved.

Working within a perturbative approach, Zwanziger's idea is to impose the condition of positive definiteness on the sum of all 
the lowest-lying eigenvalues of the Faddeev-Popov operator. Specifically, the Faddeev-Popov operator is written as 
\begin{eqnarray}
\mathcal{M}^{ab}  ~=~ \mathcal{M}^{ab}_{0} + \mathcal{M}^{ab}_{1} 
~=~ -\delta^{ab}\p^{2} + gf^{abc}A^{c}_{\mu}\p_{\mu}\;.
\end{eqnarray}
Then, after all the lowest-lying eigenvalues\footnote{Remember that 
the trivial
lowest eigenvalue is not taken into account.} associated to the ``non-perturbed'' 
operator
$\mathcal{M}^{ab}_{0}$ have been identified, the eigenvalue equation for the full (or ``perturbed'') 
operator is
written in matrix notation as
\begin{eqnarray}
\mathcal{M}S ~=~ S\Lambda_{\min} 
\end{eqnarray}
where $\mathcal{M}$ stands for the full FP operator; $S$ is the matrix 
composed 
by the
eigenstates of $\mathcal{M}$ lying on the columns, related to the lowest lying 
eigenvalues; and $\Lambda_{\min}$ stands for the diagonal matrix of the lowest lying eigenvalues 
of
$\mathcal{M}$.

The {eigenstates} of $S$ and $\Lambda_{\min}$ are treated perturbatively
with
respect to the coupling constant. To be specific
\begin{eqnarray}
S ~=~ \sum_{n=0}^{\infty} S_{n}
\qquad \text{and} \qquad
\Lambda_{\min} ~=~ \sum_{n=0}^{\infty} \Lambda_{n}
\;,
\end{eqnarray}
where $S_{0}$ and $\Lambda_{0}$ stand for eigenstates corresponding to the lowest lying 
eigenvalue and 
to the ``unperturbed FP operator'' $\mathcal{M}_{0}$, respectively. Hence, the zero-order 
term reads
\begin{eqnarray}
\mathcal{M}_{0}S_{0} ~=~ S_{0}\Lambda_{0} 
\;.
\end{eqnarray}
Using the orthogonality condition applied to each subspace, 
$\mathcal{H}_{n}$ generated 
by the eigenstates of $S_{n}$ with respect to the zero order subspace 
$\mathcal{H}_{0}$,
Zwanziger was able to solve the eigenvalue equation at each order.

When the infinite volume limit is taken, some simplifications occur so that a 
general expression
for the lowest lying eigenvalues can be derived. Then, Zwanziger substituted the 
stronger
condition -- ``the sum of all lowest lying eigenvalues shall be positive'' -- by the weaker
condition -- ``the trace of the sum must be positive''. Therefore, with the 
general
expression for the eigenvalues, one can derive the trace, obtaining
\begin{eqnarray}
\Tr\Lambda ~=~ 
2\left( \frac{2\pi}{L} \right)^{2}
\Bigg(
d(N^{2}-1)
- \frac{1}{V} \int d^{4}x d^{4}y \; 
g^{2}f^{abc}f^{adl}\,A^{b}_{\mu}(x)
\left[ \mathcal{M}^{-1} \right]^{cl} A^{d}_{\mu}(y) \delta(x-y)
\Bigg) > 0
\;.
\end{eqnarray}
Thus, the condition should be implemented in the partition function as
\begin{eqnarray}
Z_{GZ} ~=~ \int \mathcal{D}\phi \; \theta(dV(N^{2}-1) - H(A)) \e^{-(S_{YM} + 
S_{gf})}
\;,
\label{gzpartition}
\end{eqnarray}
with
\begin{eqnarray}
H(A) ~=~ 
\frac{1}{V} \int d^{4}x d^{4}y \; 
g^{2}f^{abc}f^{adl}\,A^{b}_{\mu}(x)
\left[ \mathcal{M}^{-1} \right]^{cl} A^{d}_{\mu}(y) \delta(x-y)
\label{horizonfunction}
\end{eqnarray}
where \eqref{horizonfunction} is called the horizon function. 

The partition function \eqref{gzpartition} represents a \textit{uniform 
ensemble}, where only
gauge field configurations that satisfy the condition $H(A) \leq dV(N^{2}-1)$ are {included}. In other words, it
assigns non-zero probability to physical configurations whose energy lies 
within a specific
range; otherwise,
it 
assigns zero
probability. 

Making use of a geometric result that a volume limited by a hyper-surface, such as $H(A) < dV(N^{2}-1)$, becomes concentrated on the limiting hypersurface as the number of dimensions increases. This can be understood through the following simple example: Consider an $n$-sphere. When computing its volume, we essentially integrate the volume element $r^n dr d\Omega_{n-1}$, where $\Omega_{n-1}$ parametrizes the $n-1$ angular variables. Now it is obvious that, for large $n$, the integral will get most of its contributions from the highest values of $r$, i.e.~the ones near the radius of the hyper-sphere, as $r^n$ is much higher there. Therefore, it is not difficult to see that, in the thermodynamic limit, the \textit{uniform ensemble} becomes a 
\textit{microcanonical ensemble}. The partition function of a microcanonical ensemble reads
\begin{eqnarray}
Z_{GZ} ~=~ \int \mathcal{D}\phi \; \delta(dV(N^{2}-1) - H(A)) \e^{-(S_{YM} + 
S_{gf})}
\;.
\end{eqnarray}
That is, only gauge field configurations satisfying the condition $H(A) = 
dV(N^{2}-1)$ are assigned non-zero probability. This condition is called the \textit{horizon 
condition}.

The integral representation of {the} $\delta$-function leads us to
\begin{eqnarray}
Z_{GZ} ~=~ 
\int_{-\infty + i\varepsilon}^{\infty + i\varepsilon} \frac{d\beta}{2\pi i}
\e^{-f(\beta)}
\;,
\label{hfdlkfj}
\end{eqnarray}
with $f(\beta) ~=~ - \ln \mathcal{W}(\beta)$ and
\begin{eqnarray}
\mathcal{W}(\beta) ~=~ 
\int \mathcal{D}\phi \; 
\e^{- \ln \beta}
\e^{-\left[S_{YM} + S_{gf} + \beta \left(H(A) - dV(N^{2}-1)\right) \right]}
\;.
\end{eqnarray}
Let us make use of the saddle point approximation to compute the 
integral
\eqref{hfdlkfj}. The necessary condition to use the saddle point 
approximation is given by the following equation:
\begin{eqnarray}
\frac{d f(\beta)}{d\beta}\Bigg\vert_{\beta = \beta^{\ast}} ~=~ 0 
\;{.}
\label{gapeq0}
\end{eqnarray}
Once this condition is satisfied, the approximation becomes exact in 
the infinite
volume limit. Namely, 
\begin{eqnarray}
Z_{GZ} ~=~ 
\e^{-f(\beta^{\ast})}
\;.
\end{eqnarray}
The necessary condition \eqref{gapeq0} for the saddle point is called the \textit{gap 
equation}.

Finally, the partition function becomes
\begin{eqnarray}
Z_{GZ} ~=~ 
\e^{-f(\beta^{\ast})}
\;,
\end{eqnarray}
{which describes} a \textit{canonical ensemble}, or Boltzmann ensemble, in the 
thermodynamic limit.

From now on in this paper the Gribov parameter $\beta$ will be replaced by 
$\gamma^{4}$ for simplicity and to keep track of the mass dimension of the Gribov parameter. 
In particular,  the Gribov parameter of the Lorentz symmetric YM theory will be denoted $\gamma_{0}^{4}$.

\subsection{The gap equation}

As a consequence of the Gribov restriction, a non-local mass term for the gauge 
field is introduced into the action, accounting for non-perturbative effects. Fortunately, such a non-local GZ term can be rewritten in a local form \cite{Zwanziger:1989mf}. Namely, the localized Gribov-Zwanziger action reads \cite{Dudal:2009bf,Vandersickel:2012tz},
\begin{equation} 
S_{GZ} = S_{YM} + S_{gf} + S_0+S_\gamma  \;, 
\label{sgz2}
\end{equation}
with
\begin{equation}
S_0 =\int d^{4}x \left( {\bar \varphi}^{ac}_{\mu} (\partial_\nu D^{ab}_{\nu} )
\varphi^{bc}_{\mu} - {\bar \omega}^{ac}_{\mu}  (\partial_\nu D^{ab}_{\nu} ) 
\omega^{bc}_{\mu}
- gf^{amb} (\partial_\nu  {\bar \omega}^{ac}_{\mu} ) (D^{mp}_{\nu}c^p) 
\varphi^{bc}_{\mu}
\right) \;, 
\label{s0}
\end{equation}
{\bf where $D^{ab}_{\nu}=\partial_\nu\delta^{ab} +gf^{acb}A_\nu^c$ is the covariant derivative in the adjoint representation of the $SU(N)$ group} and 
\begin{equation}
S_{\gamma_{0}} =\; \gamma_{0}^{2} \int d^{4}x \left( 
gf^{abc}A^{a}_{\mu}(\varphi^{bc}_{\mu} + {\bar \varphi}^{bc}_{\mu})\right)-4 
\gamma_{0}^4V (N^2-1)
\;.
\label{hfl}
\end{equation} 
In the actions $S_{0}$ and $S_{\gamma}$, $\bar{\phi}_{\mu}^{ab}$ and $\phi_{\mu}^{ab}$ are bosonic auxiliary fields, while $\bar{\omega}_{\mu}^{ab}$ and $\omega_{\mu}^{ab}$ are fermionic auxiliary fields. Together they are called the Gribov ghosts.

Within perturbation theory, the gap equation (\textit{i.e.} equation 
\eqref{gapeq0}) and the
gluon two point function can be explicitly computed at tree-level. To that end, 
it is sufficient to consider only terms of the local action \eqref{sgz2} that are either quadratic in the quantum fields or constant. Performing a Fourier transformation one ends up with
\begin{equation}
Z_{GZ}^{\text{quad}} ~=~ \int [\d A] \; \left[ \det -\p^{2} \right]
\exp \left\{-\frac{1}{2}  \int \frac{\d^{d}q}{(2\pi )^{d}}  \;  
A_{\mu}^{a}(q)
K_{\mu\nu}^{ab}
A_{\nu}^{b}(-q) 
- 4V\gamma^{\ast 4}(N^{2}-1)
\right\}  
\;.
\label{Zq0}
\end{equation}
with
\begin{eqnarray}
K_{\mu\nu}^{ab} ~=~ \delta^{ab}
\left[  
\left( q^{2} +  \frac{2Ng^{2}\gamma_{0}^{\ast4}}{q^{2}} \right)\delta_{\mu \nu}   
+ \left( \frac{1}{\Delta} -1 \right) q_{\mu }q_{\nu }
\right]
\label{dfdflkjd}
\end{eqnarray}
Note that in \eqref{Zq0}, the FP ghosts and the Gribov
ghosts were integrated out; the 
parameter $\Delta$ stands for the FP gauge fixing parameter and if 
$\Delta \to 0$, the Landau gauge is recovered. The Gribov parameter 
$\gamma_{0}^{\ast 4}$ represents
the solution of the gap equation \eqref{gapeq0} so that in the thermodynamic
limit the saddle point approximation becomes exact, leading us to
\begin{eqnarray}
Z_{GZ}^{\text{quad}} ~=~ 
\e^{-f(\gamma_{0}^{\ast})}
\;.
\end{eqnarray}

After some algebraic manipulations, one can derive the following 
expression for
$f(\gamma_{0})$:
\begin{eqnarray}
f(\gamma_{0}) &=& 4\gamma_{0}^{4}V(N^{2}-1) 
- \ln \gamma_{0}^{4}
- \frac{3V(N^2-1)}{4}
\int \frac{d^4 p}{(2\pi)^4}
\ln\left(p^2+ \frac{2\gamma_{0}^{4} Ng^2}{p^2}\right).
\label{gluonpropagator}
\end{eqnarray}
From the saddle point method condition 
\[
\frac{df(\gamma_{0})}{d\gamma_{0}^{2}}\Bigg\vert_{\gamma_{0}^{2} = \gamma_{0}^{\ast 2}} ~=~ 0
\;,
\]
and in the thermodynamic limit, the explicit expression for the 
gap equation is given by:
\begin{equation}
1 ~=~ \frac{3Ng^2}{8}\int \frac{d^{4}p}{(2\pi)^4}\frac{1}{p^4 + 2\gamma_{0}^{\ast 
4} 
Ng^{2}}.
\label{gapequation}
\end{equation}

Therefore, it must be clear that the gap equation \eqref{gapequation} has to be 
solved, so that the Yang-Mills theory makes sense. Such a condition comes from 
the 
necessary condition \eqref{gapeq0} for a saddle point used in the thermodynamic 
limit to implement the Gribov restriction. In \cite{Capri:2014jqa}, the explicit solution for \eqref{gapequation} was found, and it reads
\begin{eqnarray}
\gamma^{2}_{0} = \bar{\mu}^{2} \e^{\frac{1}{3}-\frac{64\pi^{2}}{3Ng^{2}}} 
\,.
\label{gribovsolution}
\end{eqnarray}
Due to the asymptotic freedom of the YM theory, the perturbative approach can only be applied in the regime of large momenta. As was pointed out in \cite{Gribov:1977wm}, the Gribov parameter can only be accessed in the regime of small momenta. This scenario can be pictured from \eqref{gribovsolution}. It can be seen that, for the large momentum regime, $\gamma_{0}^{2}$ becomes irrelevant, unlike at small momenta. This non-perturbative feature is also seen in the gauge propagator. Finding the inverse of \eqref{dfdflkjd} and taking the Landau limit $\Delta \to 0$, the gauge propagator reads
\begin{equation}
\langle A_\mu^a(k)A_\nu^b(-k)\rangle ~=~ 
\delta^{ab}\frac{k^2}{k^4+\gamma_{0}^{\ast \, 4}}
\left(\delta_{\mu\nu}-\frac{k_\mu k_\nu}{k^2}\right)
\,,
\end{equation}
which can be rewritten as
\begin{equation}
\langle A_\mu^a(k)A_\nu^b(-k)\rangle ~=~ 
\delta^{ab}
\frac{1}{2}\left(\frac{1}{k^2+i\gamma_{0}^{\ast \, 2}}+\frac{1}{k^2-i\gamma_{0}^{\ast \, 2}}\right)
\left(\delta_{\mu\nu}-\frac{k_\mu k_\nu}{k^2}\right)
 \,.
\label{gribovpropagator}
\end{equation}
From \eqref{gribovpropagator}, it is clear that after Gribov's restriction, the gluon propagator
displays complex conjugate poles, which prevents {assigning} an asymptotic single particle interpretation to it, in
the sense that its K\"all\'en-Lehmann representation is not always positive, meaning that the rotation
from Euclidean space to Minkowski space is not well-defined  \cite{Sorella:2010it}.

Since our effective Lorentz-breaking YM theory, including the CPT-even 
aether term, still displays gauge freedom, it is interesting and important to study the 
effects of such an aether term on the gap equation in the Landau gauge.

\section{Yang-Mills-Gribov-Zwanziger-aether action quantization}
As we have just seen, a CPT-even coupling term (\textit{i.e.} the 
aether-like term) 
arises from radiative corrections to the 1PI two-point function of the gauge 
field. This was shown by 
considering Lorentz violation and CPT-odd coupling terms between the gauge and 
the 
fermionic fields in Yang-Mills theories. Thus, it seems reasonable to investigate the 
influence of
such an aether-like term on the gauge field propagator at tree level. 
For this, we will consider an effective model, where the aether-like term is 
present introducing a
Lorentz-breaking Yang-Mills theory within the Landau gauge. The YM-aether non-Abelian action reads
\begin{equation}
S^{\text{Mink}}=\int d^4x 
\left(-\frac{1}{4}\left(F_{\mu\nu}^a\right)^2-\frac{\alpha}{2}a^\mu 
F^a_{\mu\nu}a_\delta F^{\delta\nu a}\right),
\label{ymcfjactionmink}
\end{equation}
where $a_\rho$ is the constant Lorentz-breaking vector, which is dimensionless 
in four dimensions. The $\alpha$ parameter is equal to the constant $\frac{\kappa e^2}{3\pi^2}$ in \eqref{aether}, while $a_{\mu}=\frac{b_{\mu}}{m}$, if we suggest that the aether term was generated as a quantum correction. This parameter plays a key role in our study and will be examined in the next section. 
The Euclidean action reads,
\begin{eqnarray}
S&=&\int d^4x \left(\frac{1}{4}\left(F_{\mu\nu}^a\right)^2+\frac{\alpha}{2}a_\mu F_{\mu\nu}^a a_\delta F_{\delta\nu}^a\right).
\label{ymlvaction}
\end{eqnarray}
From now on we work in the Landau gauge. Following the procedure 
described in the Section \ref{gribovsection}, the quadratic part of \eqref{ymlvaction} reads
\begin{equation}
S=\int 
\frac{d^4k}{(2\pi)^4}\left(\frac{1}{2}\tilde{A}_\mu^a(k)Q_{\mu\nu}^{ab}
\tilde{A}_\nu^b(-k)\right),
\label{actionprop}
\end{equation}
where 
\begin{align}
Q_{\mu\nu}^{ab}&=&\delta^{ab}\left[\left(k^2+\frac{\gamma^4}{k^2}\right)\delta_{\mu\nu}+\left(\frac{1}{\Delta}-1\right)k_\mu k_\nu+{\alpha}\left((a\cdot k)^2 \delta_{\mu\nu}-(a\cdot k) a_\nu k_\mu - a_\mu k_\nu (a\cdot k)+k^2 a_\mu a_\nu\right)\right]
\label{opq}
\end{align}
where $\gamma^4=\frac{\beta Ng^2}{2V(N^2-1)}$ is known as the Gribov parameter and $\Delta$ is the parameter responsible for the Landau gauge fixing.

\subsection{The gap equation}
\label{gapequationaether}
In this section, we compute the gap equation in the presence of an aether term 
based on the steps presented in Section \ref{gribovsection}. Starting 
with \eqref{ymlvaction}, the gluon propagator, as in \eqref{gluonpropagator}, 
reads
\begin{equation}
\langle A_\nu^a(k)A_\gamma^b(p)\rangle=\delta(p+k)\mathcal{N}\int 
\frac{d\beta}{2i\pi\beta}e^{\beta}(\det Q_{\nu\gamma}^{ab})^{-1/2} 
(Q_{\nu\gamma}^{ab})^{-1}.
\end{equation}
Computing the determinant of \eqref{opq}, we find,
\begin{equation}
(\det Q_{\nu\gamma}^{ab})^{-1/2}=e^{\frac{-1}{2}\det\ln 
Q_{\nu\gamma}^{ab}}=e^{\frac{-1}{2}\Tr\ln Q_{\nu\gamma}^{ab}}
\;.
\label{detq}
\end{equation}
We now compute $\tr\ln Q_{\mu\nu}^{ab}$. For any diagonalizable matrix $\mx M$ with eigenvalues $\lambda_i$, we have that
\begin{equation}
	\tr \ln \mx M = \sum_i \ln\lambda_i \;.
\end{equation}
For any vector $v_\mu$ orthogonal to both $k_\mu$ and $a_\mu$, we have
\begin{equation}
	Q_{\mu\nu}^{ab} v_\nu = \delta^{ab} \left( k^2+\frac{\gamma^4}{k^2} + \alpha (a\cdot k)^2\right) v_\mu \;,
\end{equation}
which gives us the first $d-2$ eigenvalues. Besides that, for $k_\mu$ we have:
\begin{equation}
	Q_{\mu\nu}^{ab} k_\nu = \delta^{ab} \left( \frac{\gamma^4}{k^2} + \frac1\Delta k^2\right) k_\mu \;,
\end{equation}
As a result we have one more eigenvalue. In order to find the last eigenvalue, we consider the vector in the $a_\mu$ and $k_\mu$ plane orthogonal to $k_\mu$:
\begin{equation}
	Q_{\mu\nu}^{ab} \left(a_\nu-\frac{a\cdot k}{k^2} k_\nu\right) = \delta^{ab} \left( k^2+\frac{\gamma^4}{k^2} + \alpha k^2a^2\right) \left(a_\mu-\frac{a\cdot k}{k^2} k_\mu\right) \;.
\end{equation}
Finally we have:
\begin{eqnarray}
&&	\tr\ln Q_{\mu\nu}^{ab} = (N^2-1) \left[ (d-2)\sum_k \ln\left(k^2+\frac{\gamma^4}{k^2} + \alpha (a\cdot k)^2\right) + \sum_k\ln\left(\frac{\gamma^4}{k^2} + \frac1\Delta k^2\right) +\right.\nonumber\\&+&\left.\sum_k \ln\left(k^2+\frac{\gamma^4}{k^2} + \alpha k^2a^2\right) \right]\nonumber\\
&=& (N^2-1) \left[ (d-2)V\int \frac{d^dk}{(2\pi)^d} \ln\left(\xi(\theta)k^4+{\gamma^4}\right) +V\zeta(a)^{d/4}\int \frac{d^dq}{(2\pi)^d} \ln\left( q^4+{\gamma^4}\right) \right] \;,
\end{eqnarray}
where we have used that $\int d^dk\ln k^2$ is zero within the framework of dimensional regularization and taken the Landau gauge limit $\Delta\to0$. Also, we have defined $\zeta(a)=1+\alpha a^2$, $\xi(\theta)=1+\alpha a^2\cos^2\theta$, $q^2=\sqrt{\zeta}k^2$ and used the fact that $(a\cdot k)^2=a^2k^2\cos^2\theta$. Therefore, from \eqref{detq}, we have that
\begin{eqnarray}
(\det Q_{\nu\gamma}^{ab})^{-1/2}&=&\exp\left[-\frac{(N^2-1)}{2}\right.\times\nonumber\\&\times&\left.
\left( (d-2)V\int \frac{d^dk}{(2\pi)^d} \ln\left(\xi(\theta)k^4+{\gamma^4}\right) +V\zeta(a)^{d/4}\int \frac{d^dq}{(2\pi)^d} \ln\left( q^4+{\gamma^4}\right) \right)\right],
\end{eqnarray}
so, the new version of \eqref{gluonpropagator} reads\footnote{Here the 
Gribov parameter has been redefined as $\gamma^{4} ~=~ \frac{\beta 
Ng^2}{N^2-1}\frac{2}{dV}$.}
\begin{eqnarray*}
f(\beta)&=&\beta-\ln\beta\nonumber\\
&-&\frac{(N^2-1)}{2}\left( (d-2)V\int \frac{d^dk}{(2\pi)^d} \ln\left(\xi(\theta)k^4+\frac{\beta 
Ng^2}{N^2-1}\frac{2}{dV}\right) +\right.\nonumber\\ &+&\left.
V\zeta(a)^{d/4}\int \frac{d^dq}{(2\pi)^d} \ln\left( q^4+\frac{\beta 
Ng^2}{N^2-1}\frac{2}{dV}\right) \right).
\end{eqnarray*}
In the thermodynamic limit, the saddle point approximation condition for $\beta$ requires $f'(\beta_0)=0$, where $\beta_0$ is the value of $\beta$ that 
minimizes the vacuum energy. Thus,
\begin{eqnarray}
0=1-\frac{1}{\beta_0}
-{Ng^2}\left( \frac{(d-2)}{d}\int \frac{d^dk}{(2\pi)^d} \frac{1}{\xi(\theta)k^4+\frac{\beta_0 
Ng^2}{N^2-1}\frac{2}{dV}} +\frac{\zeta(a)^{d/4}}{d}\int \frac{d^dq}{(2\pi)^d}\frac{1}{q^4+\frac{\beta_0 
Ng^2}{N^2-1}\frac{2}{dV}} \right).
\end{eqnarray}
The $1/\beta_0$ term can be neglected\footnote{The spacetime volume is 
infinite: $V\sim\infty$. If we set $\beta_0\sim V$ we keep the term finite and 
non zero.} and we obtain,
\begin{eqnarray}
d={Ng^2}\left( (d-2)\int \frac{d^dk}{(2\pi)^d} \frac{1}{\xi(\theta)k^4+\frac{\beta_0 
Ng^2}{N^2-1}\frac{2}{dV}} +{\zeta(a)^{d/4}}\int \frac{d^dq}{(2\pi)^d}\frac{1}{q^4+\frac{\beta_0 
Ng^2}{N^2-1}\frac{2}{dV}} \right).
\label{gapeqint}
\end{eqnarray}

Let us now look at each of the integrals of \eqref{gapeqint}. The first one
needs to be treated carefully, since its integrand is a function of the angular variable {$\theta$}. After some algebraic manipulations and
computational work, which are detailed in Appendix \ref{appendix}, one can derive the following expression for this first integral, within the
$\overline{\text{MS}}$ renormalization scheme,
\begin{align}
(d-2)\int \frac{d^dk}{(2\pi)^d} \frac{1}{\xi(\theta)k^4+\frac{\beta_0 
Ng^2}{N^2-1}\frac{2}{dV}}
&=
\frac{-4(\sqrt{1+a^{2}\alpha}-1)}{a^{2}\alpha (4\pi)^{2}}\ln\left( \frac{\gamma^{2}}{\bar{\mu}^{2}} \right) 
-\frac{2}{(4\pi)^2 }{}_{2}F_{1}^{(0,0,1,0)} \left(\frac12,1,2;-a^{2}\alpha \right) 
\nonumber\\
 &
- \frac{1}{(4\pi)^2} \, {}_{2}F_{1}^{(0,1,0,0)} \left(\frac12,1,2;-a^{2}\alpha \right)
\,.
\label{wighwigh2}
\end{align}
The explicit expression of $\frac{2}{(4\pi)^2 }{}_{2}F_{1}^{(0,0,1,0)} \left(\frac12,1,2;-a^{2}\alpha \right)$
and $\frac{2}{(4\pi)^2 }{}_{2}F_{1}^{(0,1,0,0)} \left(\frac12,1,2;-a^{2}\alpha
\right)$ can also be found at the very end of the Appendix \ref{appendix}.

The second integral of \eqref{gapeqint}, in its turn, can be easily computed
within $\overline{\text{MS}}$, and reads
\begin{eqnarray}
{\zeta(a)^{(1-\frac{\epsilon}{4})}}\int \frac{d^dq}{(2\pi)^d}\frac{1}{q^4+\frac{\beta_0 Ng^2}{N^2-1}\frac{2}{dV}}
=
\frac{\zeta(a)}{(4\pi)^2}
\left(1-\ln\frac{\gamma^2}{\bar{\mu}^2}+\frac{1}{2}\ln\zeta(a)\right)
\,.
\label{wighwigh}
\end{eqnarray}
In both expressions \eqref{wighwigh2} and \eqref{wighwigh}, $\bar{\mu}^{2}$ accounts
for the renormalization mass parameter of the system according to the
$\overline{\text{MS}}$ scheme.

Collecting the renormalized expression of each integral in \eqref{gapeqint}, one can write down the equation that describes the behavior of the Gribov parameter. Namely, 
\begin{align}
&
\frac{\gamma^{2}}{\bar{\mu}^{2}}
=
\nonumber \\
&
\exp\left\{
\frac{
-a^{2}\alpha 
\left[
\frac{4(4\pi)^{2}}{N g^{2}} 
-\zeta(a) (1 + \frac12 \ln \zeta(a))
+2 {}_{2}F_{1}^{(0,0,1,0)}(\frac12,1,2;-a^{2}\alpha)
+{}_{2}F_{1}^{(0,1,0,0)}(\frac12,1,2;-a^{2}\alpha)
\right]
}
{
4(\sqrt{1+a^{2}\alpha} -1) + a^{2}\alpha\zeta(a)
}
\right\}.
\label{owihgoig}
\end{align}

Using Wolfram Mathematica, equation \eqref{owihgoig} can be rewritten in the following manner:
\begin{eqnarray}
\gamma^{2} = \bar{\mu}^{2} \e^{\frac{1}{3}-\frac{64\pi^{2}}{3Ng^{2}}} +
\mathcal{O}(\alpha)
\,.
\label{ewhoehbq}
\end{eqnarray}
Therefore, in the very special limit $\alpha \to 0$, {\it i.e.} when the effective
Lorentz symmetry breaking disappears, Gribov's usual Yang-Mills Lorentz symmetric result in \eqref{gribovsolution} is recovered. 
It is quite clear from equation
\eqref{ewhoehbq} that the smaller the magnitude of the LSB parameter $a^{2}\alpha$, the closer the Gribov parameter $\gamma^{2}$ is {to} the usual
Lorentz symmetric $\gamma_{0}^{2}$ given by equation \eqref{gribovsolution}.
This feature is evident in the third plot (on the bottom) of Figure \ref{plotsofgamma}.

In the Figure \ref{plotsofgamma}, the behavior of the Gribov parameter on \eqref{owihgoig} is presented
in three plots. Notice that for typical values of $a^{2}\alpha$ \cite{Colladay:2018jic,Kostelecky:2008ts},
which is much smaller than 1, and within the regime of sufficiently small coupling constant $g < 1$, $\gamma^{2}$ goes to zero {exponentially}. For instance, in the
top right plot of Figure \ref{plotsofgamma}, one can clearly see that the smaller the coupling constant,
the less relevant is the Gribov parameter $\gamma^{2}$. 
In the bottom plot of this same Figure we have four curves: three of them for different values of the LSB parameter identified by the ``dashed'' curve for $a^{2}\alpha = 1$, the ``dotted'' curve for $a^{2}\alpha = 0.5$, and the ``dot-dashed'' curve for $a^{2}\alpha = 0.3$; and the fourth curve that is shown with a solid line representing the usual Lorentz symmetric Gribov parameter given by equation \eqref{gribovsolution}. Two important pieces of information can easily be read off from this bottom plot: first, the smaller the LSB parameter, the closer the curve is to the usual Lorentz symmetric Gribov parameter, $\gamma_{0}^{2}$, as is already known from equation \eqref{ewhoehbq}; and second, that the existence of a LSB aether-like term makes the Gribov parameter more relevant at smaller values of the coupling constant (comparing with the behavior of $\gamma_{0}^{2}$). Considering the asymptotic freedom behavior of this theory, one may conclude that, when considering the existence of the aether-like Lorentz symmetry violating operator, the Gribov parameter becomes more relevant at lower energy scales, when comparing with the Lorentz symmetric scenario.

From the top left and bottom plots of Figure \ref{plotsofgamma}, 
one can verify that, for typical values of $a^{2}\alpha$ (which is much smaller 
than $10^{-1}$), the Gribov parameter becomes relevant only for considerably high 
values of $g$ ($\gg 1$). In other words, for small enough values of the coupling 
constant, $\gamma^{2}$ goes to zero exponentially for any value of the LSB parameter, 
recovering, then, the UV features of the theory.

Graphical analyses of $\gamma^{2}/\gamma^{2}_{0}$ -- here, $\gamma^{2}$ is given by \eqref{owihgoig} and $\gamma_{0}^{2}$
denotes the usual, Lorentz symmetric, Gribov parameter given by 
\eqref{gribovsolution} -- were conducted as well to see the effect of the LSB
parameter, and lead us to the same conclusion. In Figure \ref{gammaLSB.gamma_vs_g}, this was {done} by keeping $a^{2}\alpha$ fixed to investigate how the effect of the LSB varies with $g$, while in Figure \ref{gammaLSB.gamma_vs_alpha} we kept the coupling constant {fixed} and let $a^{2}\alpha$ vary {freely}. From both these figures it is clear that {$\gamma^{2}/\gamma_{0}^{2} ~\to~ 1$ at stronger coupling regimes, and that {this} limit is reached faster for smaller magnitudes of the LSB parameter.
We can also see that how {the} smaller $a^{2}\alpha${, i.e. the aether term, is}, {the} faster $\gamma^{2}$ coincides with $\gamma^{2}_{0}$. In other words, the aether term will have {a} strong influence in the non-perturbative regime, if it is big enough.

\begin{figure}[h]
\begin{center}
\includegraphics[width=.45\textwidth]{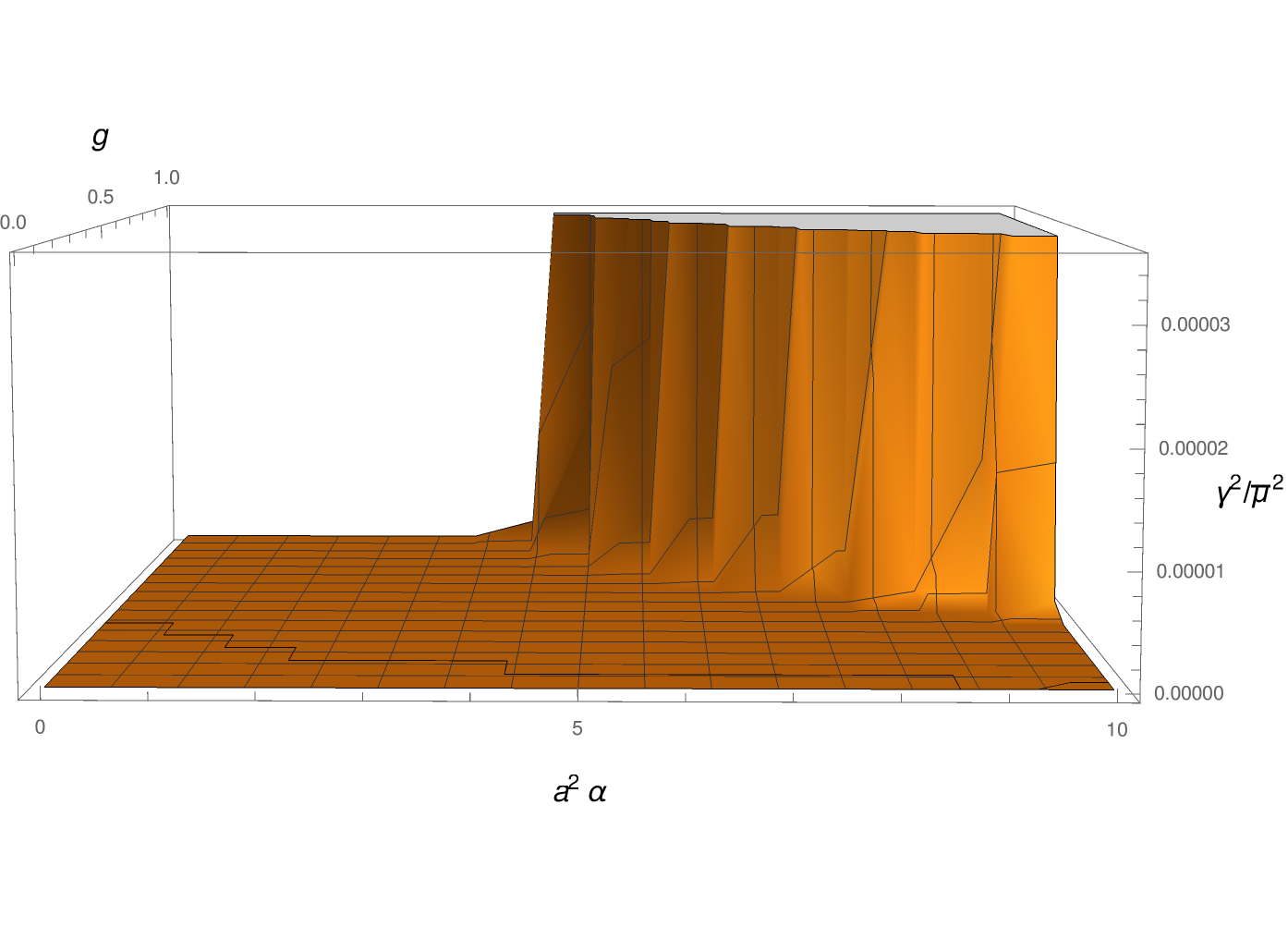} 
\hspace{10mm}
\includegraphics[width=.45\textwidth]{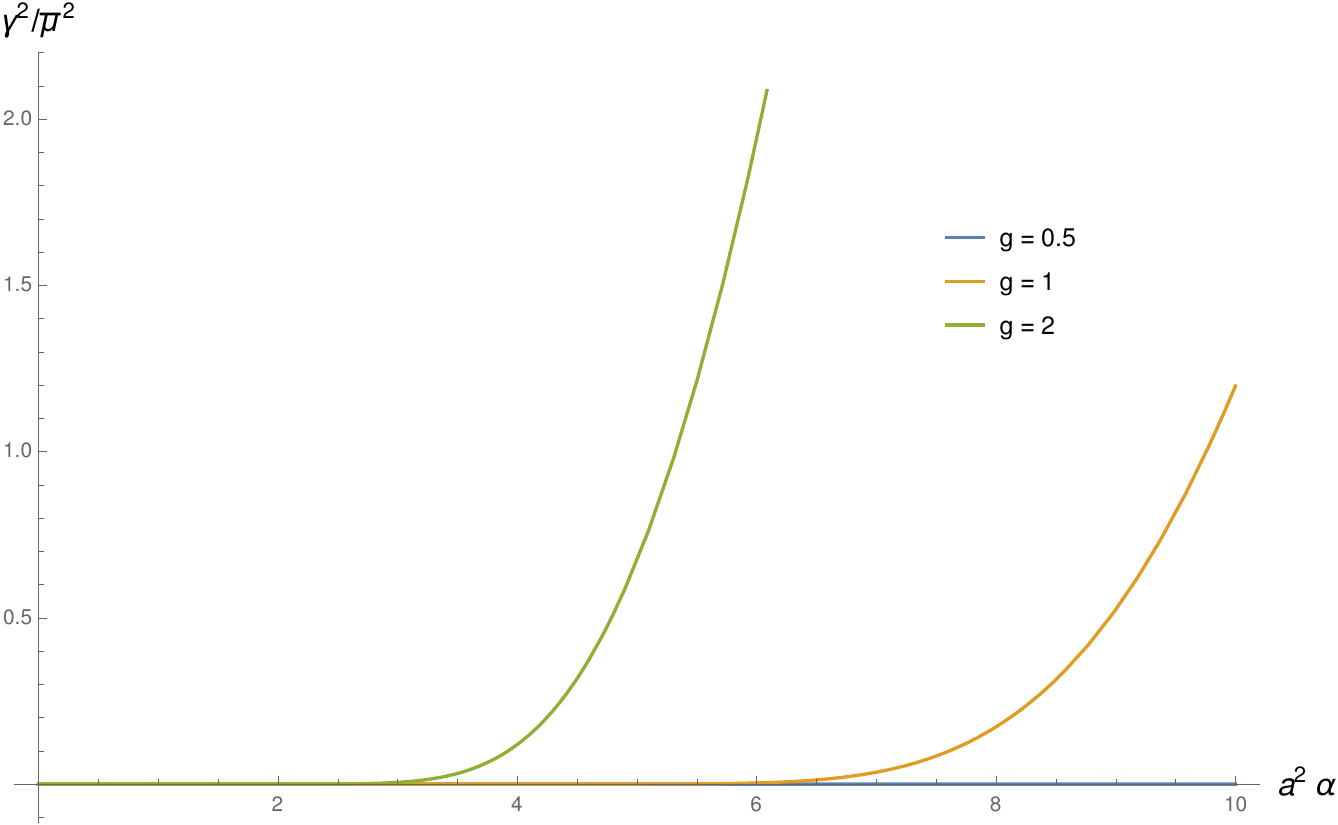}
\includegraphics[width=.5\textwidth]{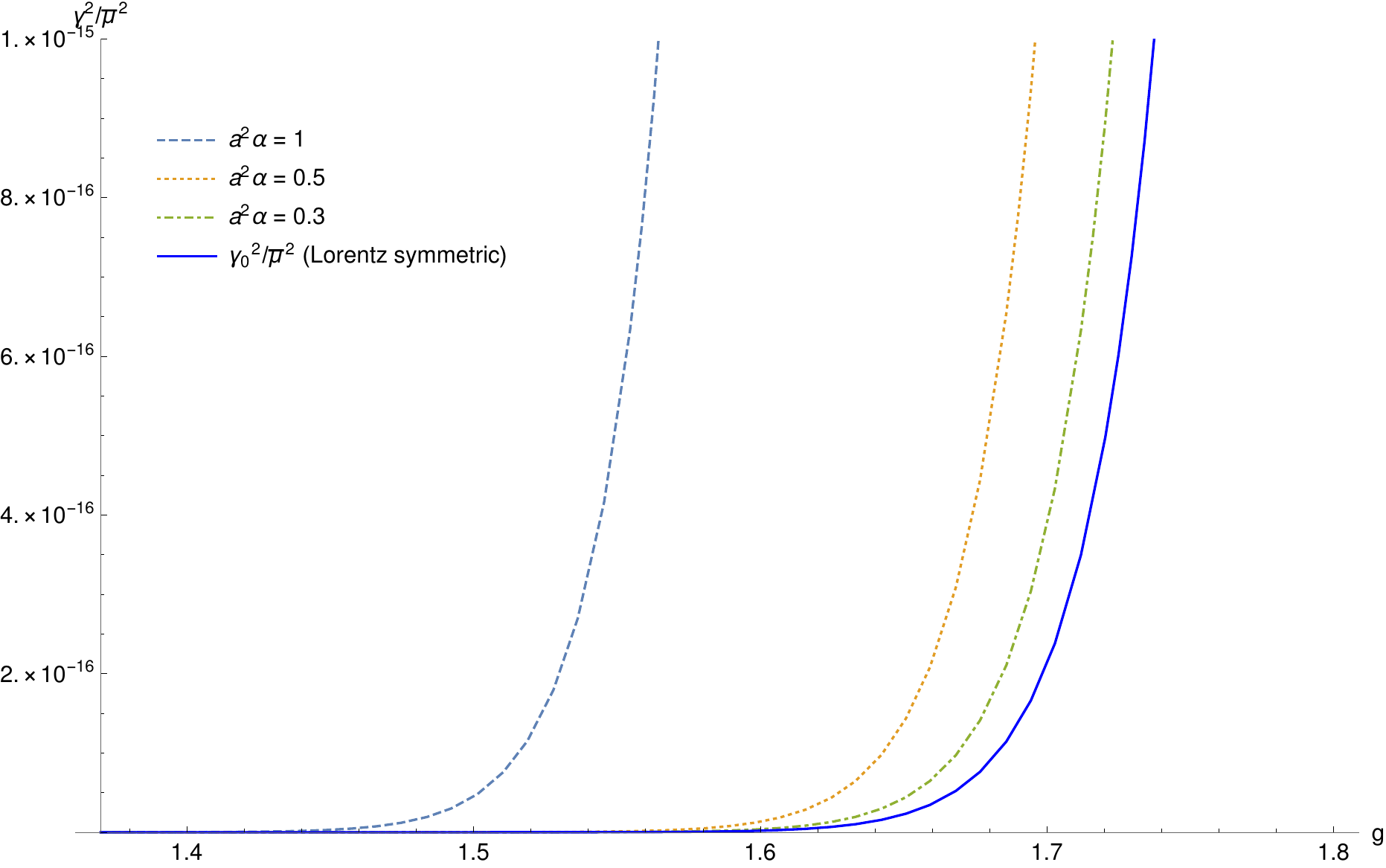}
\end{center}
\caption{Top left: 3D plot of $\gamma^{2}/\bar{\mu}^{2}$ with $g$ in the interval
$[0,1]$ and $a^{a}\alpha$ in $[0,10]$; Top right: three plots of
$\gamma^{2}/\bar{\mu}^{2} ~{\mathrm{vs}~} a^{2}\alpha$, one for $g = 0.5$ represented by {a} blue line
(almost indistinguishable from the horizontal axis), one for $g = 1$ represented by {an} orange line, and one for $g=2$ represented by {a} green line{.} Bottom: four plots of the Gribov parameter (in {units} of the $\MSbar$ renormalization parameter), $\gamma^{2}/\bar{\mu}^{2}$, {first} three for different values of $a^{2}\alpha$ and the last one for the Lorentz symmetric $\gamma_{0}^{2}/\bar{\mu}^{2}$ scenario.
}
\label{plotsofgamma}
\end{figure}

\begin{figure}[h]
\begin{center}
\includegraphics[width=.6\textwidth]{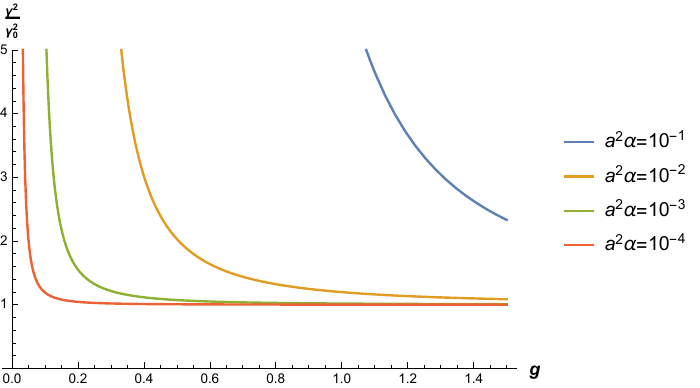}
\end{center}
\caption{
$\gamma^{2}/\gamma^{2}_{0}$ vs $g$ keeping $a^{2}\alpha$ fixed, where $\gamma^{2}$ is given by \eqref{owihgoig} and $\gamma_{0}^{2}$
denoting the usual, Lorentz symmetric, Gribov parameter given by 
\eqref{gribovsolution}.
}
\label{gammaLSB.gamma_vs_g}
\end{figure}

\begin{figure}[h]
\begin{center}
\includegraphics[width=.6\textwidth]{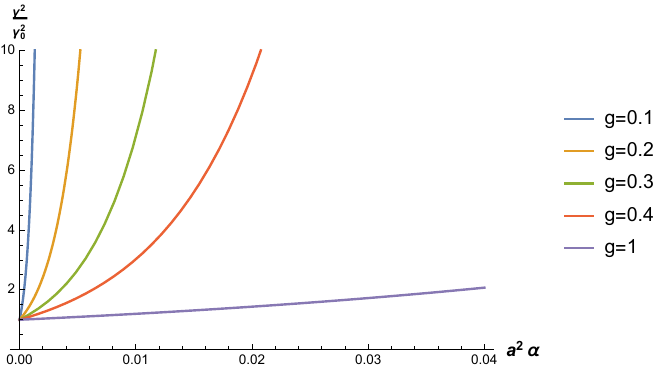}
\end{center}
\caption{
$\gamma^{2}/\gamma^{2}_{0}$ vs $a^{2}\alpha$ keeping $g$ fixed, where $\gamma^{2}$ is given by \eqref{owihgoig} and $\gamma_{0}^{2}$
denoting the usual, Lorentz symmetric, Gribov parameter given by 
\eqref{gribovsolution}.
}
\label{gammaLSB.gamma_vs_alpha}
\end{figure}

\subsection{The gauge propagator}
As {was} explained before {in} Section \ref{gzquatization}, the gauge field propagator changes {as a consequence of the GZ framework}: the poles of the propagator acquire complex mass values. As we are no longer working with pure Yang-Mills, in this section we compute the gauge field propagator and show how its poles are affected by the procedure. From \eqref{actionprop}, we have that gauge propagator reads
\begin{eqnarray}
\langle A_\mu^a(k)A_\nu^b(k)\rangle
&
=
&
\delta^{ab}F(k)\left[\left(\delta_{\mu\nu}-\frac{k_{\mu}k_{\nu}}{k^2}\right)-\left(\frac{\alpha(a\cdot k)^2}{\left(k^2+\frac{\gamma^4}{k^2}+{\alpha}k^2 a^2\right)}\right)\frac{k_\mu k_\nu}{k^2}+\right.\nonumber\\ 
&
+
&\left.
\left(\frac{\alpha(a\cdot k)}{\left(k^2+\frac{\gamma^4}{k^2}+{\alpha}k^2 a^2\right)}\right)a_\nu k_\mu+\left(\frac{\alpha (a\cdot k)}{\left(k^2+\frac{\gamma^4}{k^2}+\alpha a^2k^2\right)}\right)a_\mu k_\nu-\right.\nonumber\\
&
-
&\left.
\left(\frac{\alpha k^2}{\left(k^2+\frac{\gamma^4}{k^2}+\alpha a^2k^2\right)}\right)a_\mu a_\nu\right]
\label{aetherpropagator}
\end{eqnarray}
where
\begin{equation}
F(k)=\frac{1}{\left(k^2+\frac{\gamma^4}{k^2}+\alpha(a\cdot k)^2\right)}.
\label{f}
\end{equation}
From the definition of $F(k)$ in \eqref{f}, the IR behavior of the YM propagator found in \eqref{gribovpropagator} {can be seen to remain unchanged} also here. In order to make this explicit, $F(k)$ can be rewritten in the following way
\begin{equation}
F(k)=\frac{k^2}{\left(\xi(\theta)k^4+{\gamma^4}\right)}=\frac{1}{2}\left(\frac{1}{\sqrt{\xi(\theta)}k^2+i\gamma^{\ast \, 2}}+\frac{1}{\sqrt{\xi(\theta)}k^2-i\gamma^{\ast \, 2}}\right)
\end{equation}
As was pointed out after \eqref{gribovpropagator}, one of the implications is {that} we cannot rotate back to Minkowski space. In the previous section, we showed that, in the regime where $\alpha a^2$ lives{,} the Gribov problem becomes irrelevant. Thus, a rotation to Minkowski space can be performed and the gauge propagator reads:
\begin{eqnarray}
\langle A_\mu^a(k)A_\nu^b(k)\rangle
&
=
&
\frac{\delta^{ab}}{k^2+\alpha(a\cdot k)^2}\left[\delta_{\mu\nu}-\frac{k_{\mu}k_{\nu}}{k^2}-\frac{\alpha(a\cdot k)^2}{k^2+{\alpha}a^2 k^2}~\frac{k_\mu k_\nu}{k^2}+\right.\nonumber\\ 
&
+
&\left.
\frac{\alpha(a\cdot k)}{k^2+{\alpha}a^2 k^2}~a_\nu k_\mu+\frac{\alpha (a\cdot k)}{k^2+\alpha a^2k^2}~a_\mu k_\nu-\frac{\alpha k^2}{k^2+\alpha a^2k^2}~a_\mu a_\nu\right].
\end{eqnarray}
Some remarks about this propagator {must be} presented. We can easily check that it is transversal as it must be. It displays two types of dispersion relation: first, the usual one $k^2=0$, i.e.~in Minkowski space, $E^2=\vec{k}^2$, and the new one $k^2+\alpha(a\cdot k)^2=0$ which must be treated separately for the cases of time-like and space-like $a_{\mu}$. Let us briefly discuss these cases.

If $a_{\mu}$ is time-like, $a_{\mu}=(a,0,0,0)$, we have $E^2-\vec{k}^2+\alpha a^2E^2=0$, {such that} the relation $E^2>\vec{k}^2$ necessary for {the} absence of tachyons is {realized} if $\alpha<0$.

If $a_{\mu}$ is space-like, e.g. $a_{\mu}=(0,a,0,0)$, we have $E^2=\vec{k}^2-\alpha a^2k^2_x$, so, again the relation $E^2>\vec{k}^2$ is {realized} if $\alpha<0$.

However, we note that in principle there is no fundamental restrictions on the sign of $\alpha$ while, within our scheme of perturbative generation, it turns out to be negative as is required by the dispersion relations.

\section{Conclusion}
In this paper we have treated {the proper path integral quantization of the YM-aether} theory
in the Landau gauge. We proved that such an effective CPT-even Lorentz-breaking YM theory 
can be consistently quantized in the Landau gauge according to the 
Gribov-Zwanziger quantization prescription. In other words, following the GZ 
approach to get rid of the Gribov copies in the Landau gauge, we proved that 
the aether-like term does not spoil the consistency condition, \textit{i.e.} 
the gap equation exists and can still be solved within our effective theory 
displaying CPT-even Lorentz breaking. A whole analysis about different regimes of the theory was carried out in order to realize the impact of the aether coupling on different regimes of the theory. We have shown that the Lorentz symmetry breaking can influence the non-perturbative regime of the theory for a case {whether} the aether coupling has a big values.

Finally, it would be interesting to study the renormalizability of the 
effective 
YM-aether theory, within the Algebraic Renormalization 
prescription,  and to restudy this procedure {with the inclusion of} a Higgs field to see 
its impact on the gap equation and on the {the poles of the gluon propagator}.

\section*{Acknowledgments}
We thank D.~Bazeia, F.~Canfora, A.~Giacomini and R.~F.~Sobreiro for fruitful 
discussions. C.~P.~{F.} {was} supported by Ministry of Science and Technology (MoST) project 1087636. I.~F.~J. {was} supported by CAPES, project $88887.357904/2019-00$.
The work by A. Yu. P. {was} partially supported by the CNPq, project 301562/2019-9.

\appendix
\section{The gap equation integrals}
\label{appendix}

In this appendix we provide some details concerning the first integral of the
gap equation, given {in} equation \eqref{gapeqint}. {That is namely} the integral
whose integrand depends on the $\theta$ direction, 
\begin{equation}
\int \frac{d^dk}{(2\pi)^d} \frac{1}{\xi(\theta)k^4+\frac{\beta_0 
Ng^2}{N^2-1}\frac{2}{dV}}
\,,
\end{equation}
with $\xi(\theta) = 1 + a^{2}\alpha \cos\theta$.

{R}edefining the {direction of the} momentum $k${,} it is possible to rewrite {this} integral as
\begin{equation}
\int \frac{d^dk}{(2\pi)^d}
\frac{1}{\xi(\theta)^{d/4}} 
\frac{1}{k^4+\frac{\beta_0 Ng^2}{N^2-1}\frac{2}{dV}}
\end{equation}

Introducing spherical coordinates in $d$ dimensions (for dimensional
regularization) and setting up the background constant field $a_{\mu}$ parallel
to the first direction $\hat{e}_{1}$, such that $a\cdot k = ak\cos\theta_{1}$
and, therefore, $\xi(\theta) = \xi(\theta_{1})$, we
get
\begin{equation}
	\frac1{(2\pi)^d} 
\int_0^\infty dk \ k^{d-1} 
\int_0^\pi d\theta_1 \ \sin^{d-2}\theta_1 
\int_{0}^{\pi} 
\cdots
\int_{0}^{2\pi}d\theta_{d-1} 
\;
\frac{1}{\xi(\theta_{1})^{d/4}} 
\frac{1}{k^4+\frac{\beta_0 Ng^2}{N^2-1}\frac{2}{dV}}
\,.
\end{equation}

Notice that it is possible to rewrite {this} integral as
\begin{equation}
	\frac{\int_0^\pi\frac{\sin^{d-2}\theta_1}{(1+\alpha a^2\cos^2\theta_{1})^{d/4}}
d\theta_{1}}{\int_0^\pi \sin^{d-2}\theta_1 d\theta_{1}} 
\int \frac{d^dk}{(2\pi)^d} 
\frac{1}{k^4+\frac{\beta_0 Ng^2}{N^2-1}\frac{2}{dV}}
\,.
\label{lhgeihg}
\end{equation}

In its turn, the numerator of the global factor of equation \eqref{lhgeihg} can
be expanded in powers of $\alpha$ leading us to
\begin{equation}
	\sum_{n=0}^\infty (\alpha a^2)^n \begin{pmatrix}-d/4\\n\end{pmatrix}
\int_0^\pi \sin^{d-2} \cos^{2n}\theta d\theta \;,
\label{numowihrg}
\end{equation}
{where the} integral can be performed by noticing that 
\begin{equation}
	\frac d{d\theta} \sin^{d-1}\theta \cos^{2n-1}\theta = (2n+d-2) \sin^{d-2}\theta\cos^{2n}\theta - (2n-1) \sin^{d-2}\theta\cos^{2n-2}\theta \;,
\end{equation}
such that
\begin{equation}
	\int_0^\pi \sin^{d-2} \cos^{2n}\theta d\theta = 0 + \frac{2n-1}{2n+d-2} \int_0^\pi \sin^{d-2}\theta \cos^{2n-2}\theta d\theta
\end{equation}
where we assumed $d > 1$. {As such, the} integral can be {computed to give}
\begin{equation}
	\int_0^\pi \sin^{d-2} \cos^{2n}\theta d\theta = \frac{(\frac12)_n}{(\frac d2)_n} \int_0^\pi \sin^{d-2}\theta d\theta \;.
\end{equation}
Here and further, we use the Pochhammer symbols given by
\begin{equation}
(x)_n=\frac{\Gamma(x+1)}{\Gamma(x-n+1)}  \,.
\end{equation}

Then, the numerator \eqref{numowihrg} can be recast as 
\begin{equation}
	\sum_{n=0}^\infty (\alpha a^2)^n \begin{pmatrix}-d/4\\n\end{pmatrix} \frac{(\frac12)_n}{(\frac d2)_n} \;.
\label{series}
\end{equation}
Since
\begin{equation}
	\begin{pmatrix}-d/4\\n\end{pmatrix} = (-1)^n \begin{pmatrix}d/4+n-1\\n\end{pmatrix} = (-1) \frac{\Gamma(\frac d4+n)}{\Gamma(\frac d4) n!} = (-1) \frac{(\frac d4)_n}{n!} \;,
\label{nb}
\end{equation}
we {the} get
\begin{equation}
	\sum_{n=0}^\infty \frac{(-\alpha a^2)^n}{n!} \frac{(\frac12)_n(\frac d4)_n}{(\frac d2)_n} = {}_2 F_1\left(\tfrac12,\tfrac d4;\tfrac d2;-\alpha a^2\right) \;.
\end{equation}
Thus, finally, the original integral can be written in terms of the usual
{Gribov} gap equation {type} integral multiplied by the hypergeometric function
${}_2 F_1\left(\tfrac12,\tfrac d4;\tfrac d2;-\alpha a^2\right)$, namely,
\begin{equation}
\int \frac{d^dk}{(2\pi)^d} \frac{1}{\xi(\theta)k^4+\frac{\beta_0 
Ng^2}{N^2-1}\frac{2}{dV}}
= {}_2 F_1\left(\tfrac12,\tfrac d4;\tfrac d2;-\alpha a^2\right) 
\int \frac{d^dk}{(2\pi)^d} 
\frac{1}{k^4+\frac{\beta_0 Ng^2}{N^2-1}\frac{2}{dV}}
\,.
\end{equation}

According to Mathematica, the hypergeometric function ${}_{2}F_{1}\left(\frac12,\frac{(4-\epsilon)}{4},\frac{(4-\epsilon)}{2};-a^{2}\alpha \right)$ can be expanded around $\epsilon= 4 -d$, for $\epsilon << 1$, {to give}
\begin{align}
{}_{2}F_{1} \left(\frac12,\frac{(4-\epsilon)}{4},\frac{(4-\epsilon)}{2};-a^{2}\alpha \right) 
&= 
\frac{2(\sqrt{1+a^{2}\alpha}-1)}{a^{2}\alpha} - \frac{\epsilon}{2} \, {}_{2}F_{1}^{(0,0,1,0)} \left(\frac12,1,2;-a^{2}\alpha \right) 
\nonumber \\
&
- \frac{\epsilon}{4} \, {}_{2}F_{1}^{(0,1,0,0)} \left(\frac12,1,2;-a^{2}\alpha \right) 
+ {\cal O}(\epsilon^{2})
\end{align}
where ${}_{2}F_{1}^{(0,0,1,0)} \left(\frac12,1,2;-a^{2}\alpha \right) $ stands for the derivative of  ${}_{2}F_{1} \left(\frac12,\frac{(4-\epsilon)}{4},\frac{(4-\epsilon)}{2};-a^{2}\alpha \right)$ w.r.t. $\epsilon$ in the third argument of the hypergeometric function and then taken at $\epsilon =0$. {Below} one can find an explicit expression of these terms.

The integral that usually appears in the gap equation within dimensional regularization is
\begin{align}
(d-2)\bar{\mu}^{4-d} \, \int \frac{d^{d}k}{(2\pi)^{d}} \, f(k^{4} + \gamma^{4}) &= 
(d-2)\bar{\mu}^{4-d} \, \int \frac{d^{d}k}{(2\pi)^{d}} \, \frac{1}{k^{4} + \gamma^{4}}
\quad \text{with} \quad d=4-\epsilon
\nonumber \\
& =
\frac{2}{(4\pi)^{2}} \left[
\frac{2}{\epsilon} - \gamma_{E} + \ln (4\pi) - \ln\left( \frac{\gamma^{2}}{\bar{\mu}^{2}} \right) + {\cal O}(\epsilon)
\right]
\,,
\end{align}
with $\bar{\mu}$ accounting for the renormalization mass scale.

Therefore, the original expression becomes
\begin{align}
&
{}_{2}F_{1} \left(\frac12,\frac{(4-\epsilon)}{4},\frac{(4-\epsilon)}{2};-a^{2}\alpha \right) 
(d-2)\int \frac{d^{d}k}{(2\pi)^{d}} \, f(k^{4} + \gamma^{4}) = 
\nonumber \\
&=
\frac{4(\sqrt{1+a^{2}\alpha}-1)}{a^{2}\alpha (4\pi)^{2}} \left[
\frac{2}{\epsilon} - \gamma_{E} + \ln (4\pi) - \ln\left( \frac{\gamma^{2}}{\bar{\mu}^{2}} \right) 
\right] -
\nonumber \\
&
- \frac{2}{(4\pi)^2 }{}_{2}F_{1}^{(0,0,1,0)} \left(\frac12,1,2;-a^{2}\alpha \right) 
- \frac{1}{(4\pi)^2} \, {}_{2}F_{1}^{(0,1,0,0)} \left(\frac12,1,2;-a^{2}\alpha \right)
+ {\cal O}(\epsilon).
\end{align}
Within {the} $\overline{\text{MS}}$ renormalization scheme, one has
\begin{align}
\label{werigh}
&
{}_{2}F_{1} \left(\frac12,\frac{(4-\epsilon)}{4},\frac{(4-\epsilon)}{2};-a^{2}\alpha \right) 
(d-2)\int \frac{d^{d}k}{(2\pi)^{d}} \, f(k^{4} + \gamma^{4}) = 
\nonumber \\
&=
\frac{4(1-\sqrt{1+a^{2}\alpha})}{a^{2}\alpha (4\pi)^{2}}\ln\left( \frac{\gamma^{2}}{\bar{\mu}^{2}} \right) 
- \frac{2}{(4\pi)^2 }{}_{2}F_{1}^{(0,0,1,0)} \left(\frac12,1,2;-a^{2}\alpha \right) 
~-
\nonumber \\
&
- \frac{1}{(4\pi)^2} \, {}_{2}F_{1}^{(0,1,0,0)} \left(\frac12,1,2;-a^{2}\alpha \right)
\,.
\end{align}

Now, let us write down the explicit {expressions} of ${}_{2}F_{1}^{(0,1,0,0)}
\left( a,b,c;z \right)$ and ${}_{2}F_{1}^{(0,0,1,0)}\left( a,b,c;z \right)$ . 
Namely, they are given by
\begin{equation}
{}_{2}F_{1}^{(0,1,0,0)} \left( a,b,c;z \right) = 
\sum_{n=0}^\infty   \frac{(a)_{n+1} (b)_{n+1}}{(c)_{n+1}}  \frac{(z)^{n+1}}{(n+1)!} \sum_{s=0}^{n}\frac{1}{s+b}
\end{equation}
and
\begin{equation}
{}_{2}F_{1}^{(0,0,1,0)} \left( a,b,c;z \right) = 
-\sum_{n=0}^\infty   \frac{(a)_{n+1} (b)_{n+1}}{(c)_{n+1}}  \frac{(z)^{n+1}}{(n+1)!} \sum_{s=0}^{n}\frac{1}{c+s}
\,.
\end{equation}

Therefore, in our case we have
\begin{equation}
{}_{2}F_{1}^{(0,1,0,0)} \left(\frac12,1,2;-a^{2}\alpha \right) = 
\sum_{n=0}^\infty   \frac{(1/2)_{n+1}}{{(n+1)}}  \frac{(-a^{2}\alpha)^{n+1}}{(n+1)!} \sum_{s=0}^{n}\frac{1}{s+1}
\end{equation}
and
\begin{equation}
{}_{2}F_{1}^{(0,0,1,0)} \left(\frac12,1,2;-a^{2}\alpha \right) = 
\sum_{n=0}^\infty   \frac{(1/2)_{n+1}}{{(n+1)}}  \frac{(-a^{2}\alpha)^{n+1}}{(n+1)!} \sum_{s=0}^{n}\frac{1}{s+2}
\;.
\end{equation}

\end{document}